\RequirePackage{amsmath}

\documentclass[5p]{elsarticle}

\usepackage{bm}
\usepackage{graphicx}
\usepackage{color}
\usepackage{amssymb}
\usepackage{bbold}
\usepackage{siunitx}
\usepackage{booktabs}
\usepackage{enumitem}

\newcommand{\reivision}[1]{{\color{blue}{#1}}}

\biboptions{sort&compress}

\journal{Physica E}

\begin{document}

\begin{frontmatter}

\title{Nanowires: A route to efficient thermoelectric devices}

\author{Francisco Dom\'inguez-Adame}
\address{GISC, Departamento de F\'{\i}sica de Materiales, Universidad
Complutense, E-28040 Madrid, Spain}

\author{Marisol Mart\'{\i}n-Gonz\'{a}lez}
\address{IMM--Instituto de Micro y Nanotecnolog\'{i}a de Madrid (CNM--CSIC), Isaac Newton 8, 
PTM, E-28760 Tres Cantos, Madrid, Spain}

\author{David S\'{a}nchez}
\address{Instituto de F\'{\i}sica Interdisciplinar y Sistemas Complejos IFISC 
(UIB-CSIC), E-07122 Palma de Mallorca, Spain}

\author{Andr\'{e}s Cantarero}
\address{Instituto de Ciencia Molecular, Universitat de Val\'{e}ncia, PO Box 22085, E-46071 Valencia, Spain}

\begin{abstract}

Miniaturization of electronic devices aims at manufacturing ever smaller products, from mesoscopic to nanoscopic sizes. This trend is challenging because the increased levels of dissipated power demands a better understanding of heat transport in small volumes. A significant amount of the consumed energy in electronics is transformed into heat and dissipated to the environment. Thermoelectric materials offer the possibility to harness dissipated energy and make devices less energy-demanding. Heat-to-electricity conversion requires materials with a strongly suppressed thermal conductivity but still high electronic conduction. Nanowires can meet nicely these two requirements because enhanced phonon scattering at the surface and defects reduces the lattice thermal conductivity while electric conductivity is not deteriorated, leading to an overall remarkable thermoelectric efficiency. Therefore, nanowires are regarded as a promising route to achieving valuable thermoelectric materials at the nanoscale. In this paper, we present an overview of key experimental and theoretical results concerning the thermoelectric properties of nanowires. The focus of this review is put on the physical mechanisms by which the efficiency of nanowires can be improved. Phonon scattering at surfaces and interfaces, enhancement of the power factor by quantum effects and topological protection of electron states to prevent the degradation of electrical conductivity in nanowires are thoroughly discussed.

\end{abstract}


\end{frontmatter}

%%%%%%%%%%%%%%%%%%%%%%%%%%%%%%%%%%%%%%%%%%%%%%%%%%%%%%%%%%%
\section{Introduction}\label{sec:intro}
%%%%%%%%%%%%%%%%%%%%%%%%%%%%%%%%%%%%%%%%%%%%%%%%%%%%%%%%%%%

The ever-increasing world energy consumption is expected to grow more than $50\%$ between $2010$ and $2040$. Energy resources to meet this demand can be classified into three categories, namely fossil fuels, nuclear resources and renewable resources~\cite{Demirbas00}. Environmental risks and deterioration, limited reserves, acid precipitation and global climate change are a few of the many disadvantages of non-renewable resources (oil, coal, gas) and nuclear power. On the contrary, renewable energy sources come either directly or indirectly from the Sun and solar energy can be straightforwardly used for \emph{heating\/} and generating \emph{electricity\/}. Moving industry and home energy needs towards environmentally friendly sources will eventually make non-renewable energy demands to dwindle.

In a recent review, He and Tritt highlighted the fact that heat and electricity are two very different forms of energy~\cite{He17}. Heat is a ubiquitous energy, but with low quality. In the opposite end, electricity is a versatile energy resource, but its production is demanding. In this context, thermoelectricity aims at finding efficient bridges between them. Converting heat emanating from a hot source such as the Sun, engines and boilers directly into electricity lies at the very heart of thermoelectric devices. Similarly, refrigerators and heat pumps without moving parts or greenhouse gasses driven by an electric current is another niche application in thermoelectricity.

From a fundamental point of view, thermoelectricity research heads towards developing new materials with high performance as well as exploring innovative physical mechanisms for heat-to-electricity conversion. The so-called dimensionless \emph{figure of merit\/} $ZT$ dictates the thermoelectric efficiency of the system~\cite{Goldsmid10}. Its definition is $ZT=\sigma S^2 T / \kappa$, where $S$ stands for the Seebeck coefficient, and $\sigma$ and $\kappa$ are the electric and thermal conductivities at a temperature $T$ \cite{Villagonzalo99}, respectively. Most researchers consider a value of $ZT$ of the order of $3$ the minimum for widespread applications. With some exceptions, however, bulk materials are well below this limit but progress has been made in the last few years. For instance, Zhao \emph{et al.} reported a value of $ZT=2.6\pm 0.3$ at $\SI{923}{\kelvin}$ in a simple layered crystalline material, SnSe~\cite{Zhao14}. But the major drawback of thermoelectric devices still is the low efficiency compared to other energy-conversion technologies. 

The unfavorable interdependence of $S$, $\sigma$ and $\kappa$ ravels the efforts to implement strategies for improvement, especially those based on non-toxic materials. It is necessary to minimize the thermal conductivity $\kappa$ while keeping the power factor $\sigma S^2$ high. Thermoelectric efficiency is constrained by the Wiedemann-Franz law in bulk metals, implying that the ratio $\kappa/\sigma T$ is independent of temperature~\cite{Franz1853}. This ratio is also known as the Lorenz number, and within a degenerate electron gas model it is calculated as $L=\kappa/\sigma T=(1/3)(\pi k_B/e)^2=\SI{2.45e-8}{\watt\ohm/\kelvin^2}$, where $k_B$ is the Boltzmann constant and $e$ is the elementary electric charge. Most common metals are found to agree with this value of $L$ within $10\%$ at room temperature. 

It is most important to bear in mind that both electrons and lattice vibrations can contribute to the heat current through a system subjected to a temperature gradient. Consequently, thermal conductivity splits as $\kappa=\kappa_\mathrm{el} + \kappa_\mathrm{ph}$. It is then apparent that reducing the lattice thermal conductivity $\kappa_\mathrm{ph}$ by increasing phonon scattering is one of the most promising pathways to achieving good thermoelectric materials. This is especially relevant in materials with low electron density, such as semiconductors and insulators, because the electron thermal conductivity $\kappa_\mathrm{el}$ is expected to be small. Tailored crystal structures can help reducing the contribution of the lattice to the thermal transport. Olvera \emph{et al.} demonstrated that $ZT$ values as high as $2.6$ at moderate temperature ($\leq \SI{850}{\kelvin}$) can be attained in Cu$_2$Se matrix upon partial dissolution of In and atomic-scale incorporation of CuInSe$_2$ inclusions~\cite{Olvera17}. This remarkable performance was attributed to the localization of Cu$^+$ ions induced by the incorporation of In into the crystal lattice, which reduce the thermal conductivity of the nanocomposites.

In a simple approach, the heat-carrying phonons can be pictured as a classical gas of particles and the kinetic theory predicts that the lattice thermal conductivity is $\kappa_\mathrm{ph}=(1/3)C_\mathrm{v}v\ell_\mathrm{ph}$, with $C_\mathrm{v}$ as the specific capacity of the phonons at constant volume, $v$ the phonon velocity and $\ell_\mathrm{ph}$ the phonon mean free path~\cite{Kittel86}. At temperatures of the order of the Debye temperature the specific capacity is essentially constant and, consequently, a decrease of the phonon mean free path leads to a lower value of the lattice thermal conductivity. Defects can reduce both contributions $\kappa_\mathrm{ph}$ and $\kappa_\mathrm{el}$ to the thermal conductivity but they have the adverse effect of decreasing the electric conductivity $\sigma$ as well. Anharmonic effects also contribute to lower the mean free path via phonon-phonon interactions. For instance, Morelli \emph{et al.} reported measurements of the thermal conductivity of high-quality crystals of the cubic I−-V−-VI$_2$ semiconductors AgSbTe$_2$ and AgBiSe$_2$~\cite{Morelli08}. In these materials heat conduction was found to be dominated by the lattice term and strong phonon-phonon interaction makes the phonon mean free path equals the lattice parameter.

The advent of nanotechnology has renewed attention on thermoelectric efficiency and material development at small scales. Miniaturization of electronic devices for novel applications has brought with it new challenges, such as excessive heating and subsequent power consumption and material degradation~\cite{Pop10}. In this scenario, thermoelectric devices can be used for cooling the system and reducing power consumption. Remarkably, it is now well established that nanometer-sized objects may exhibit thermoelectric efficiency unachievable with bulk materials~\cite{Hicks93a,Hicks93b,Khitun00,Venkata01,Harman02,Balandin03,Hochbaum08,Boukai08}. In particular, quantum effects enable going beyond the limitations stemmed from the classical Wiedemann-Franz law. For instance, nanodevices displaying sharp resonances in the electron transmission (such as Fano lineshapes) are considered ideal candidates for thermoelectricity at the nanoscale because the ratio $\sigma/\kappa_\mathrm{el}$ increases well above the Wiedemann-Franz limit~\cite{Mahan96,GomezSilva12,Zheng12,Chen12,Garcia13,Fu15,SaizBretin16,Wang16}. Nanometer-sized objects display enhanced phonon scattering due to their reduced size even in the absence of anharmonic effects or Umklapp processes. Combined with the sharp resonances in the electron transmission mentioned above, size effects in nanostructures yield high values of $ZT$.

Among a vast amount of different nanostructures with potential application in thermoelectricity, such as quantum dots, nanoribbons, nanorings and nanoparticles, nano\-wires (NWs) stand out for several reasons. A number of recently developed techniques enable the growth of NWs of various materials with different diameters, cross sections (square, hexagon, triangle), crystal orientations, dopings, surface roughness as well as core-shell structures~\cite{Wang06,Meyyappan09,Cao11,Lu14}. Precise control of the physical and chemical properties of the NWs enables the effective decoupling of electric and thermal conductivities, thus achieving $ZT$ one or two orders of magnitudes larger than in their bulk counterparts, as was shown in Si~\cite{Hochbaum08,Boukai08}, ZnO~\cite{Demchenko11}, Bi$_2$Te$_3$~\cite{Zhang12a,Zhang12b}, SiGe~\cite{Lee12}, CdS~\cite{Wang17} and SnSe~\cite{Hernandez18} NWs. The core-shell nanostructuring also contributes to lower the thermal conductivity of  Bi$_2$Te$_2$S--Bi$_2$S$_3$, Bi$_2$Te$_3$--Bi$_2$S$_3$, Ge--Si, and PbTe--PbSe NWs due to the scattering of phonons at the heterojunction (see Ref.~[\citenum{Quintana13}] and references therein).

In this review, we will concern ourselves with experimental and theoretical research in thermoelectric NWs, with an emphasis on general features independent of specific material systems. The review is organized as follows. In section~\ref{sec:ns} we discuss some major achievements concerning the improvement of the thermoelectric properties of nanostructures (NWs, nanoribbons, quantum rings) relative to their corresponding bulk counterparts. Section~\ref{sec:tinw} deals with novel topological insulators and the relevance of their protected surface states on the thermoelectric transport. In section~\ref{sec:onw} we review the peculiarities of the thermoelectricity based on organic NWs. Section~\ref{sec:gt} focuses on the various fabrication techniques available for growing NWs. Theoretical aspects and approaches commonly used in thermoelectric research will be briefly presented in section~\ref{sec:th}. Applications of NWs in thermoelectric devices is the subject of 
section~\ref{sec:app}. Finally, in section~\ref{sec:con} we will summarize and draw some general conclusions and prospective.

%%%%%%%%%%%%%%%%%%%%%%%%%%%%%%%%%%%%%%%%%%%%%%%%%%%%%%%%%%%
\section{Thermoelectric response of nanowires and nano\-ribbons}   \label{sec:ns}
%%%%%%%%%%%%%%%%%%%%%%%%%%%%%%%%%%%%%%%%%%%%%%%%%%%%%%%%%%%

Although ballistic electron transport in nanostructures pave the way for achieving more efficient thermoelectric devices~\cite{Koumoto13}, a concomitant decrease of the thermal conductivity is required. Graphene is a paradigmatic example of how nanostructering can improve thermoelectric performance. It is foreseen as a promising candidate for superseding Si in future nanoelectronics and its properties can be tailored to meet these criteria. Combined with hexagonal boron nitride, graphene can be efficiently used for on-chip cooling solutions due to enhanced power factor~\cite{Duan16}. Nevertheless, bulk graphene possesses one of the highest lattice thermal conductivities~\cite{Balandin08,Balandin11,Nika17} and its overall $ZT$ value is thus quite low~\cite{Anno17}. But the thermal conductivity $\kappa$ can be greatly reduced in graphene nanoribbons by rough edges~\cite{Savin10}, H--passivation~\cite{Hu10}, patterning ~\cite{Mazzamuto11,Li14,ZhangHS12,Chen10,Xu10} or hybrid nanostructuring schemes~\cite{Sevincli13}, as compared to bulk graphene. In bulk, the thermal conductivity of graphene reaches values about $\SI[input-quotient=:, output-quotient=\text{ -- }]{2000:4000}{\watt/\meter\kelvin}$~\cite{Pop12} while it is about an order of magnitude smaller in nanoribbons~\cite{Guo09}. 

\begin{figure}[htb]
\centerline{\includegraphics[width=0.8\columnwidth]{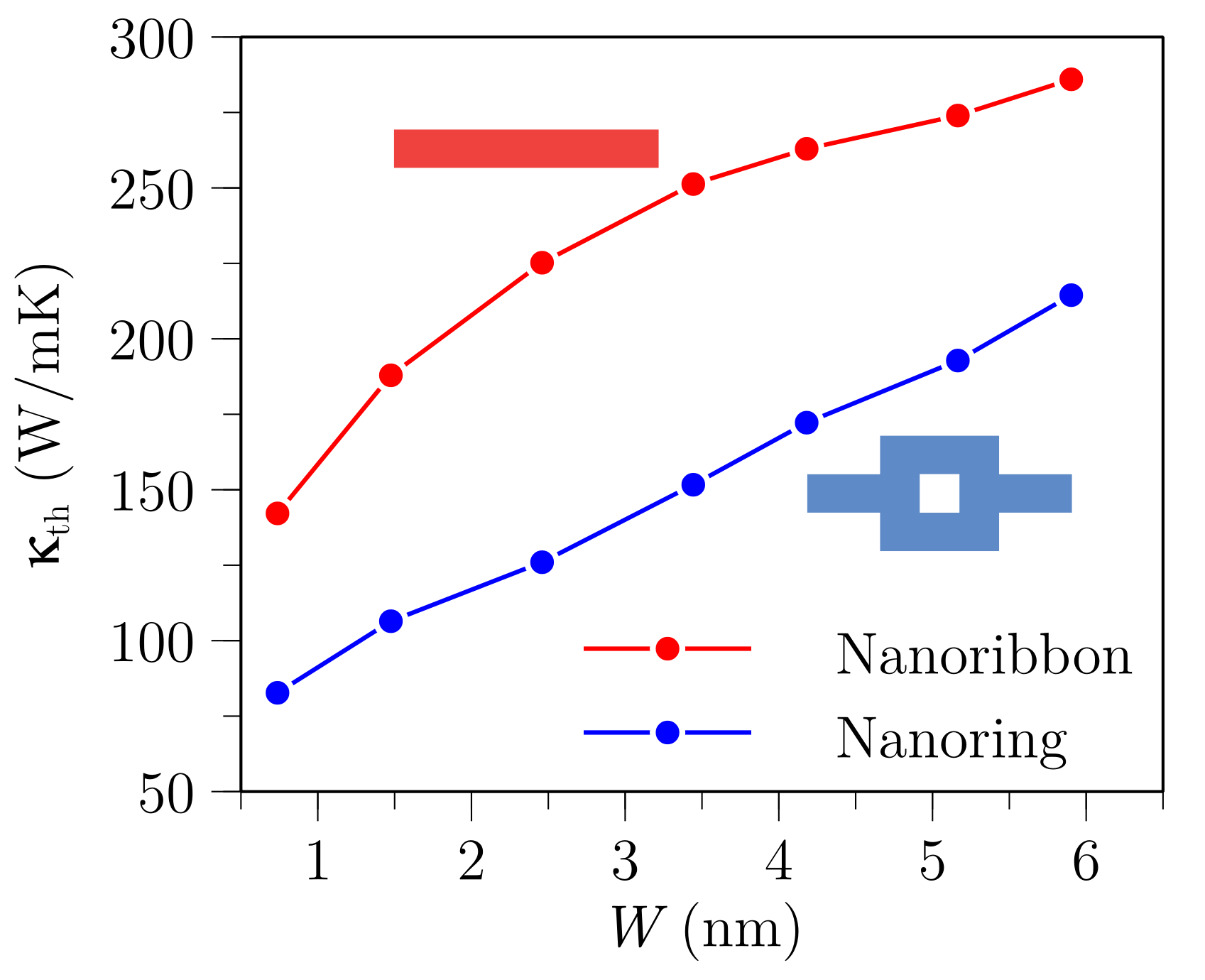}}
\caption{Lattice thermal conductivity as a function of width for armchair nanoribbons and nanorings at $T=\SI{300}{\kelvin}$. Adapted from Ref.~[\citenum{SaizBretin18}].}
\label{fig1}
\end{figure}

Additionally, non-equilibrium molecular dynamics simulations above $\SI{100}{\kelvin}$ show that narrow graphene nanorings can efficiently suppress the lattice thermal conductivity as compared to nanoribbons of the same width~\cite{SaizBretin18}. Fig.~\ref{fig1} shows the lattice thermal conductivity at room temperature of graphene nanoribbons and nanorings of widths up to $\SI{6}{\nano\meter}$. $\kappa_\mathrm{th}$ increases monotonously with $W$ in both cases, in agreement with previous results~\cite{Guo09,Evans10}. Nevertheless, the values shown in Fig.~\ref{fig1} are well below the thermal conductivity of bulk graphene. Most importantly, the occurrence of different types of edges (armchair and zig-zag) in nanorings leads to a mismatch of the vibrational modes~\cite{Mazzamuto11} and stronger scattering at the bends~\cite{Li14,Yang13}. The increased scattering of vibrational modes in nanorings results in a sizable reduction of the lattice thermal conductivity as compared to nanoribbons~\cite{SaizBretin18}. The corresponding values of $ZT$ are expected to be high due to the presence of sharp and asymmetric resonances in electron transmission that increase the Seebeck coefficient~\cite{Hicks93a,Hicks93b,Mahan96,Humphrey05,SaizBretin15}.

Semiconductors have a lower carrier density and typically the thermal conductivity is dominated by the atomic lattice. Current electronic technologies rely on Si but its thermal conductivity $\kappa\sim \SI{150}{\watt/\meter\kelvin}$~\cite{Glassbrenner64} at room temperature is too high for thermoelectric applications. Decreasing temperature yields lower values of the thermal conductivity $\kappa\sim \SI{35}{\watt/\meter\kelvin}$~\cite{Thompson61} but the power factor is reduced too. However, Si nanostructuring takes advantage of reduced-geometry effects to enhance phonon scattering while keeping electric properties almost unchanged. In this context, numerical simulations on thin-film lateral architecture of thermo-converter have recently proved that $\SI{10}{\nano\meter}$ thick Si has efficiency $10$ times larger than bulk~\cite{Haras15}. Stranz \emph{et al.} fabricated Si pillars, produced in a wafer-scale top-down process by cryogenic dry etching followed by thermal oxidation and oxide stripping, and characterized Si pillars for thermoelectric devices with diameters from $\SI{2}{\micro\meter}$ down to $\SI{170}{\nano\meter}$~\cite{Stranz11}. They found a reduction of thermal conductivity with the pillar diameter that were in agreement with a phenomenological model based on empirical results from Si NWs grown by the vapour-–liquid-–solid method. The estimated thermal conductivities for the $\SI{1000}{\nano\meter}$ and $\SI{700}{\nano\meter}$ were found to be $\SI{110}{\watt/\meter\kelvin}$ and $\SI{104}{\watt/\meter\kelvin}$, respectively. High-density and large-area vertically aligned porous Si NW arrays with different length, porosity and diameters have been successfully fabricated on the two sides of Si substrate using a simple metal-assisted chemical etching method by Zhang \emph{et al.}~\cite{Zhang15}. They reported a high ZT value of $0.493$ at room temperature, which is $77$ times higher than that measured in bulk Si ($ZT=0.0064$). Ordered dense arrays of $p$--type Si NWs were produced with a vapor-liquid-solid (VLS) method by Calaza \emph{et al.}, who estimated that $ZT$ values ranged from $0.30$ to $0.93$~\cite{Calaza15}.

\begin{figure}[htb]
\centerline{\includegraphics[width=0.9\columnwidth]{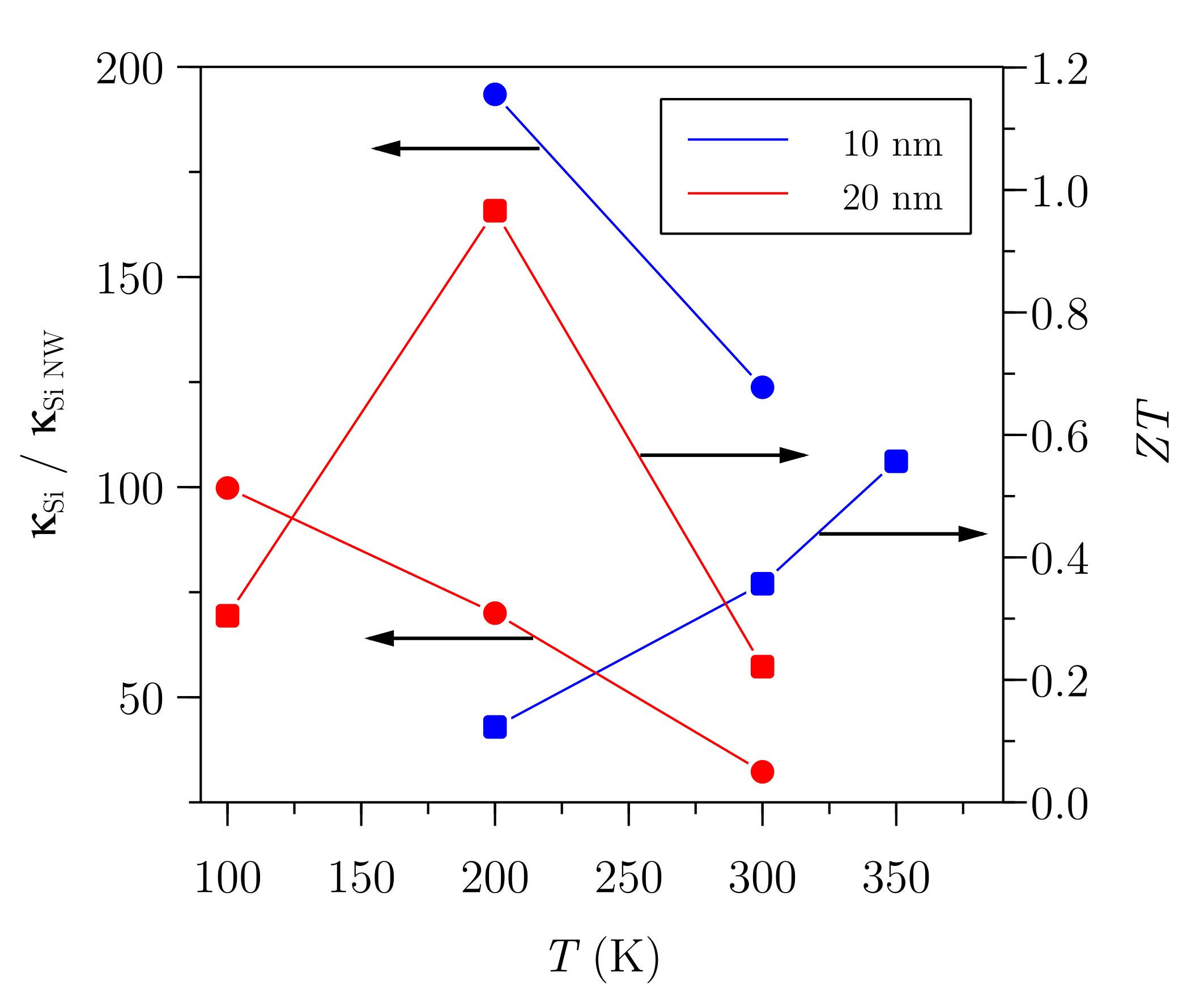}}
\caption{Temperature dependence of the thermal conductivity and $ZT$ of Si NWs of different widths, indicated in the plot. The left axis shows the thermal conductivity of Si NWs (solid circles), expressed in units of the thermal conductivity of bulk Si, and the right axis corresponds to the figure of merit (solid squares). Adapted from Ref.~[\citenum{Boukai08}].}
\label{fig2}
\end{figure}

Similarly, the introduction of large concentrations of vacancies in Si nano-films causes a $20$ fold reduction in thermal conductivity~\cite{Bennett15}. A remarkable enhancement of the thermoelectric performance of narrow Si NWs was reported by Boukai \emph{et al.}~\cite{Boukai08}. Fig.~\ref{fig2} shows the temperature dependence of the thermal conductivity and $ZT$ of Si NWs of $10$ and $\SI{20}{\nano\meter}$ width. It is observed a $200$ fold reduction in thermal conductivity and a $100$ fold enhancement in $ZT$. By independent measurements of the Seebeck coefficient, the electrical conductivity and the thermal conductivity, combined with theory, they attributed the improved efficiency to phonon scattering effects. The reduction of the thermal conductivity by nanostructuring is expected to apply to other classes of semiconductor nanomaterials~\cite{Gayner16}.

%%%%%%%%%%%%%%%%%%%%%%%%%%%%%%%%%%%%%%%%%%%%%%%%%%%%%%%%%%%
\section{Topological insulator nanowires}   \label{sec:tinw}
%%%%%%%%%%%%%%%%%%%%%%%%%%%%%%%%%%%%%%%%%%%%%%%%%%%%%%%%%%%

Topological insulators are generally defined as a class of topological materials that cannot be adiabatically connected to the vacuum without closing the insulating gap. This in turn leads to gapless modes at the boundary of these materials whilst the bulk remains an insulator~\cite{Hasan10,Hsieh12,Ando15,Bansil16,Tchoumakov17,Inhofer17}. The topological robustness is ensured as long as some discrete symmetries of the bulk are preserved, such as time-reversal symmetry~\cite{Kane10}. In three dimensions, examples of such materials include Bi$_{1-x}$Sb$_x$~\cite{Hsieh08}. Although experimentally discovered in 2008, the band gap of this alloy is rather small and it therefore limits its applications to low temperatures. A year after, a family of materials including Bi$_2$Se$_3$, Bi$_2$Te$_3$ and Sb$_2$Te$_3$ was put forward~\cite{Zhang09,Moore09}. Interestingly to this review, Bi$_2$Te$_3$ is also a well-known thermoelectric material~\cite{Kane10}. A subclass of topological insulators that rely on crystalline symmetries has also been experimentally realized in Pb$_{1-x}$Sn$_{x}$Te~\cite{Hsieh12,Assaf16,Hasan12} and Pb$_{1-x}$Sn$_{x}$Se~\cite{Dziawa12}. All these topological states of matter were already encountered in a series of seminal papers by Volkov and Pankratov in the 1980s~\cite{Volkov85,Korenman87,Agassi88,Pankratov90,Adame94,Kolesnikov97}. Remarkably, these states can be reshaped by means of external electric and magnetic fields~\cite{Diaz-Fernandez17a,Diaz-Fernandez17b,Diaz-Fernandez17c,Diaz-Fernandez18a} and doping~\cite{Bahramy12,Diaz-Fernandez18b}. Another example of these such states can be found in HgTe. However, as a result of the crystalline symmetries in HgTe, there is a degeneracy at the $\Gamma$ point and, consequently, HgTe is a semimetal. A topological insulating behaviour can be observed by lowering these symmetries. A pioneering prediction in this direction was put forward by Bernevig \emph{et al.}~\cite{Bernevig06} and later confirmed by K\"{o}nig \emph{et al.}~\cite{Konig07}. In their work, the subbands originating from a quantum well do have an energy gap. Just like in other heterostructures, as the width of the well is increased, the hole-like levels ($p$-like) move upwards in energy, whereas the electron-like levels ($s$-like) move downwards. However, differently to other heterostructures, the band-edge alignment in this material is such that the electron-like and hole-like levels can be inverted when a certain width is exceeded. As a result, there is a parity inversion above a certain critical width. It was shown by Fu and Kane~\cite{Fu07} that the system undergoes a topological phase transition where a $Z_{2}$ topological invariant changes from $\nu=0$ to $\nu=1$ due to this parity inversion. The quantum well set-up implies the observation of a two-dimensional topological insulator in HgTe. A route to have HgTe as a three-dimensional topological insulator is to apply strain on it, which can be done by using CdTe as a substrate during growth since it has a slightly different lattice constant. Nevertheless, the strain-induced energy gap is relatively small in comparison to the other materials presented above~\cite{Brune11}.

From the discussion above, it can be drawn that most topological insulator materials are based on heavy elements and display relatively narrow band gaps. The combination of these two features causes topological insulators, such as bismuth telluride (Bi$_2$Te$_3$) and selenide (Bi$_2$Se$_3$), antimony telluride (Sb$_2$Te$_3$) and tin telluride (SnTe) to perform as high-efficient thermoelectric materials at room temperature~\cite{Xu17}. In the last few years, a large number of theoretical and experimental investigations have been conducted to elucidate the mechanisms that ultimately lead to high power factors and low thermal conductivities in topological insulators. On the one side, heavy elements enhance relativistic effects and lead to narrow gaps which are favorable for room-temperature thermoelectricity~\cite{Hinsche15}. Furthermore, heavy elements are subjected to strong relativistic corrections, leading to the bulk band inversion discussed above. On the other side, heavy elements tend to reduce the lattice thermal conductivity $\kappa_\mathrm{ph}$ of the material. All these traits yield a significant improvement of the thermoelectric efficiency~\cite{Yang17}.

It is generally believed that phonon mean free paths, $\ell_\mathrm{ph}$, are much longer than those of electrons, $\ell_\mathrm{el}$. NWs were predicted long ago to exploit this difference in order to achieve a significant reduction of the lattice thermal conductivity while maintaining good electric properties~\cite{Hicks93a,Hicks93b}, provided that their diameter is larger than $\ell_\mathrm{el}$ but smaller than $\ell_\mathrm{ph}$. However, recent \emph{ab initio\/} simulations in Dirac materials ($p$-type SnTe) have shown that this is not necessarily the case due to the selective filtering of long $\ell_\mathrm{el}$ that are harmful to the Seebeck coefficient~\cite{Liu18}. Most interestingly, the robustness of the topological surface states makes them ideal candidates to prevent the detrimental effects of surface roughness and defects on the electrical transport of NWs. The relevance of surface states on electric transport was revealed by techniques based on coherent electron oscillations (Aharonov-Bohm and Shubnikov-de Haas) in Sb$_2$Te$_3$, Bi$_2$Te$_3$ and Bi$_2$Se$_3$ NWs (see Ref.~\cite{Hamdou15} and references therein).

Surface-dominated thermoelectric properties have been observed in Bi NWs by Huber \emph{et al.}~\cite{Huber11,Huber12}, although bulk Bi is classified as a trivial insulator. They reported nearly pure surface electronic transport for temperatures $T<\SI{100}{\kelvin}$ in Bi NWs with diameter around the critical value $\SI{50}{\nano\meter}$ at which the semimetal-to-semiconductor transition takes place. The presence of Dirac electrons in topological insulators enables surface-dominated transport in Bi$_{1-x}$Sb$_x$~\cite{Nikolaeva13}, Bi$_{1.5}$Sb$_{0.5}$Te$_{1.7}$Se$_{1.3}$~\cite{Hsiung15}, Bi$_2$Te$_3$~\cite{MunozRojo16}, PbSeTe~\cite{Xu15} and Bi$_2$Se$_3$~\cite{Shin16} NWs. In Sec.~\ref{sec:th} we will further discuss the origin of these surface states.

%%%%%%%%%%%%%%%%%%%%%%%%%%%%%%%%%%%%%%%%%%%%%%%%%%%%%%%%%%%
\section{Organic nanowires}   \label{sec:onw}
%%%%%%%%%%%%%%%%%%%%%%%%%%%%%%%%%%%%%%%%%%%%%%%%%%%%%%%%%%%

Organic materials, basically polymers, do not present most of the disadvantages as compared to inorganic materials. They are, in general, non toxic, there is abundance of the raw materials, they can be flexible and scalable, implemented in large areas such as facades and windows. On the other hand, they have a low thermal conductivity in bulk form, one of the major problems limiting the efficiency in bulk semiconductors in thermoelectricity. Intrinsically conducting polymers have been successfully used in the last ten years~\cite{Bharti18,Culebras14} providing a $ZT$ up to $0.4$, the same order of magnitude than that obtained with inorganic materials. However, there is no clear route to improve the efficiency of organic materials. Polymers are disordered materials with long chains and the typical theoretical models to analyze the thermal properties are phenomenological, based on the hopping of electrons from one polymer chain to another.

In principle, polymers are built or grown as disordered chains, difficulting the electronic transport~\cite{Thiessen17}. But, since these chains are one dimensional, restricting the dimension of the material, i.e. going from thin films to NWs, the chains must be ordered along the NWs and in the case of thin NWs a crystallization appears, i.e., the existence of long range order~\cite{MunozRojo14,MunozRojo15}. Actually, there are theoretical predictions~\cite{Wang11} giving a ZT$\approx 16$ for PEDOT:PSS and poly(3-hexylthiophene-2,5-diyl) (P3HT) NWs, which are referred to as \emph{molecular nanowires}. Mu\-\~noz Rojo \emph{et al}. studied the thermal conductivity of P3HT NWs embedded in anodized aluminum oxide matrix~\cite{MunozRojo14,MunozRojo15}. They have analyzed the thermal conductivity of single P3HT NWs of several diameters ($120$, $220$ and $\SI{350}{\nano\meter}$) by means of a scanning thermal microscope in a $3\omega$ configuration. The resulting thermal conductivity $\kappa$ for the NWs was found to be $\SI{0.50}{\watt\per\kelvin\per\meter}$ for the NW with $\SI{120}{\nano\meter}$ diameter, $\SI{0.70}{\watt\per\kelvin\per\meter}$ for that with $\SI{220}{\nano\meter}$ diameter and $\SI{2.29}{\watt\per\kelvin\per\meter}$ for the one with $\SI{350}{\nano\meter}$ diameter. They also measured the thermal conductivity of the composite, which decreases with decreasing diameter. They claim that the decrease is due to a poorer crystallization of the polymer inside the alumina nanochannels for the smaller diameters studied. So, the better the polymer is aligned the higher their thermal conductivity.

There are only a few works on pure polymer NWs. One interesting study was performed by Zhang \emph{et al.}~\cite{Zhang18} on PEDOT NWs doped with FeCl$_3$ (C-PEDOT) and Fe(Tos)$_3$ (T-PEDOT), deposited as a thin film.  The Seebeck coefficient was around $20.8\,\mu$V/K while the electrical conductivity was $236\,$S/m for T-PEDOT and $541\,$S/m for C-PEDOT, giving a power factor of $10$ and $23\,\mu$WK$^{-2}$m$^{-1}$, respectively and mobilities as high as $1$ and $4.5\,$cm$^2$V$^{-1}$s$^{-1}$, respectively. The enhancement of the Seebeck coefficient was explained based on X-ray photoelectron spectroscopy (XPS) measurements. Since the doping levels were similar, the fact that the Seebeck coefficient is practically the same suggests that the shape of the density of states around the Fermi level is the same. On the other hand, grazing incidence wide-angle X-ray scattering show some peaks coming from the amorphous phase, while there are some features indicating the crystallization of PEDOT when grown in the form of NWs. The position of these peaks, a smaller spacing between conjugated molecules which increases the $\pi-\pi$ interaction and an increase in the charge transfer is given as the reason for the improvement. The suggestion was confirmed from electron spin resonance measurements. Another study on PEDOT NWs is that of Taggart \emph{et al.}~\cite{Taggart11} where the NWs were grown by means of lithographic techniques on Si$_3$N$_4$-coated Si substrates. They are actually flat, with $40-90\,\mu$m thickness, $150-580\,$nm width and as long as $200\,\mu$m. They obtain a Seebeck coefficient up to $-122\,\mu$V/K and an electrical conductivity up to $40\,$S/cm. Mobilities were of the order of $9-12\,$cm$^2$V$^{-1}$s$^{-1}$. The power factor was of the order of $10^{-6}\,$W m$^{-1}$K$^{-2}$.

Most of the works developed on organic NWs are actually in polymer nanocomposites. We can distinguish two types of composites. Most of the NWs mentioned before are doped or mixed with semiconductors or metals (mostly graphene). But there is a second kind of polymer/metal-semiconducor complexes that consists actually of the immersion of semiconducting or metallic NWs in a polymer matrix, \emph{i.e.} the NWs are not the organic part. In this second class of NWs we can cite some interesting works. Abramson \emph{et al.} \cite{Abramson04}  built a thermoelectric device based on Si NWs embeded into a polymer matrix (parylene). The thermal conductivity of the composite seems to be very close to that of the perylene ($\approx 5\,$W m$^{-1}$K$^{-1}$) while they estimated a theoretical value of $1.3\,$Wm$^{-1}$K$^{-1}$. They also estimate the  figure of merit of the system, being of the order of $3.6\times 10^{-4}$.

%%%%%%%%%%%%%%%%%%%%%%%%%%%%%%%%%%%%%%%%%%%%%%%%%%%%%%%%%%%
\section{Fabrication techniques}   \label{sec:gt}
%%%%%%%%%%%%%%%%%%%%%%%%%%%%%%%%%%%%%%%%%%%%%%%%%%%%%%%%%%%

Thermoelectric NWs have been prepared by different techniques. In order to review the different approaches used to fabricate them, we divide those techniques into bottom-up (when the NW is created from smaller building blocks) and top-down techniques (when the NW is cut out from a bigger piece). Both approaches have advantages and drawbacks~\cite{Fan06}. In general, the bottom-up approaches are more advantageous than the top-down because they have a better chance of producing less defected nanostructures, with a high control over chemical composition, more homogeneity, and they can have better ordering both in the short and the long ranges. Also, when thinking about scalability, in most cases bottom-up methods are cheaper and can be produced in large areas. In the case of top-down methods, it can be claimed that they are more reproducible regarding crystallinity, doping, and composition by controlling the properties of the bulk material and that those are then transferred to the NWs once etched from the bulk. A comparison of both types of technologies is summarized in Table~\ref{table1}. As it can be observed, both approaches have advantages and disadvantages and nowadays, depending on the application and on the properties of interest, both types of methodologies can be used, being one more appropriate for certain cases. 

\begin{table*}[ht]
\caption{General comparison of NWs fabrication techniques.}
\label{table1}
\small
\centering
\setlength{\tabcolsep}{2pt}
\setlist[itemize]{leftmargin=5.0mm}
\begin{tabular}{ | m{0.12\linewidth} | m{0.25\linewidth} | m{0.48\linewidth} |} \toprule
\multicolumn{1}{|c|}{\textbf{Approach}} & 
\multicolumn{1}{|c|}{\textbf{General advantages}} & 
\multicolumn{1}{|c|}{\textbf{General disadvantages}} \\ \midrule
\multicolumn{1}{|c|}{\textbf{Top-down}} &
\begin{itemize}
\item Well-developed techniques.
\item Well known microfabrication tools.
\end{itemize} &
\begin{itemize}
\item High cost.
\item Not quick to manufacture.
\item Limited to \textquotedblleft countable\textquotedblright number of structures, though this number may be billions and billions.
\item Not suitable for large-scale production.
\end{itemize} \\ \midrule
\multicolumn{1}{|c|}{\textbf{Bottom-up}} &
\begin{itemize}
\item Able to build smaller struc\-tures than most top-down approaches.
\item Large-scale fabrication.
\item Low cost.
\end{itemize} &
\begin{itemize}
\item Require compatible surfaces or molecules.
\item Fewer tools to manipulate molecules and atoms.
\item Developments needed in precise positioning of the NWs with respect to contacts and connections, 
\item Improvements for better electrical connections are required.
\end{itemize} \\ \bottomrule
\end{tabular}
\end{table*}

A summary of some of the bottom-up and top-down fabrication methods applied to the manufacture of thermoelectric NWs are compiled below.

\subsection{Bottom-up thermoelectric nanowire fabrication}

\subsubsection{Nanowires grown from vapor phase}

Several routes for the fabrication of thermoelectric NWs have been developed in the past based on their grown from metal nanoparticles. The most commonly used is the chemical vapor deposition~(CVD) method to produce high-quality Si NWs using metal nanoparticles as catalysts, being the mechanism used the vapor--liquid--solid~(VLS) growth~\cite{Wagner64,Westwater97}. This method was initially invented in Bell Labs in 1964 by Wagner and Ellis~\cite{Wagner64} and it has been widely investigated to produce Si, Ge, and Si/Ge NWs by several groups~\cite{Morales98,Hu99,Cui00,Cui01a,Cui01b,Samuelson04,Hochbaum05,Stelzner06,Kayes07,Wu02,Li03(a),Li03(b),Hochbaum08, Tang10,Andrews11,Lim12}. The VLS mechanism consists of exploiting the catalytic properties of metal nanoparticles deposited on Si substrates. Then, at high temperatures (normally between $300$ and $\SI{800}{\celsius}$ depending on the precursor) these nanoparticles on Si are exposed to a vapor phase precursor of the NW material, like silane (SiH$_{4}$) and/or a germane (GeH$_{4}$), which forms a eutectic with the metal. Under certain conditions, the supersaturation of the silicon/metal or the germanium/metal alloy is produced and the Si and/or Ge crystallization is produced under each metal nanoparticle. Therefore, a unidirectional NW growth progresses under the nanoparticle. In this method, the selection of the appropriate precursor has been shown to be crucial, while the size of the metal particle and the growth temperature determine the NW diameter and the crystallinity of the NW. The doping, in this case, can be obtained by adding PH$_{3}$ and B$_{2}$H$_{6}$ gases in the VLS chamber, for example. In such a way, the Si NWs can be $n$ or $p$ doped, respectively. Nevertheless, it should be taken into account that the addition of these doping gases can modify the directional growth of NWs with a diameter smaller than $\SI{50}{\nano\meter}$~\cite{Fasoli12,Wallentin11}. Some of the main advantages of this growth method are its compatibility with Si technology and the facility to control doping composition and to make segmented NWs or core/shell. Finally, this is one of the few bottom-up methods that has been able to obtain the integration of arrays NWs into actual devices in a horizontal fashion and within trenches~\cite{Chaudhry07,Davila12}. 

Apart from Si and Ge NWs, the VLS method from the gaseous phase has also been used to grow wires of III-IV compounds. However, it has been less applied to grow chalcogenides NWs, although it has been extended to prepare more complex thermoelectric materials in the form of NWs, like zinc oxide, indium selenide, gallium nitride, gallium arsenide, indium phosphide, Bi$_{2}$Te$_{3}$~\cite{Hamdou13} or even topological insulator states like Bi$_{2}$Te$_{3}$~\cite{Park16} or Bi$_{2}$Se$_{3}$~\cite{Kong10}.

A variation to the VLS methods is the vapor-solid-solid (VSS). In this method the NWs are produced at lower temperature than in VLS, which makes it more suitable for the production of devices since it is more compatible with standard industrial processes. As an example, Si/Ge hetero-junction NWs with abrupt interfaces have been synthesized by VSS~\cite{Chou12}.

Another classical method that has been frequently used to grow high-quality heterostructures is molecular beam epitaxy (MBE). This method, when it is performed on a substrate with metal impurities, can generate semiconducting NWs similarly to the way VLS does~\cite{Werner06, Zakharov06} The main difference is that in MBE the metallic impurities do not act as a dissociating agent for the precursor, but promote the NWs growth~\cite{Schubert04, Werner06,Spirkoska08}. This method provides high control over the crystallinity and doping of the NWs, although it has a slow growth rate. By this method, several semiconductor NWs have been prepared, but only a few of them (like Bi$_{4}$Te$_{3}$~\cite{Wang09a}) have been studied as thermoelectric.

\subsubsection{Nanowires grown from solution}

These types of techniques have the advantage that they can produce large quantities when compared to other methods. Examples of these fabrications methods are:

\begin{enumerate}

\item The hydrothermal/solvothermal synthesis, which can produce NWs in a high-pressure solution-based environment. The NWs are produced at high rates. This technique is suitable for thermoelectric materials with high anisotropy which exhibit a preferential growth direction, such as Bi$_{2}$Te$_{3}$~\cite{Zhang12}, SbTe~\cite{Xia03}, PbTe~\cite{Tai08,Yan08}. If the synthesis is done in water, then it is called hydrothermal and if it is done in other solvents, it is referred to as solvothermal.

\item Heterostructured Bi$_{2}$Se$_{3}$ NWs~\cite{Qiu04} and PbTe nano\-rods~\cite{Qiu05} have been shown to be possible by sonoelectrochemical processes in aqueous solution at room temperature.

\item  Si and Ge NWs have also been grown by thermolysis employing a supercritical fluid-liquid-solid growth method~\cite{Heitsch11, Hanrath03} using metal nanoparticles as seeds \cite{Rackauskas09} (as it was made in the previous section). In such a way, when Si and Ge organometallic precursors are supplied into the chamber filled with a supercritical organic solvent, like toluene, NWs are grown.

\item  Sb$_{2}$Se$_{3}$ NWs were grown by using the so-called polyol synthesis method~\cite{Chen05} based on the synthesis of metallic nanostructures in high-boiling temperature alcohols. 
\end{enumerate}

Finally, we will discuss the method that is probably the most widely used to grow thermoelectric NWs: growth inside a template. We will review the kinds of templates more used as well as the fabrication methods of the NWs within them. To begin with, there are two main templates that have been used to grow thermoelectric NWs: ion-track membranes and anodic porous alumina. 

On the one hand, ion track membranes are prepared by irradiating a thick polycarbonate foil with a high-energetic ion beam. Each ion generates a damage track along its trajectory. Then, the tracks are selectively etched by introducing the membrane into a concentrated NaOH solution at $\SI{50}{\celsius}$. This produces cylindrical nanochannels randomly distributed, whose diameter increases linearly with the etching time. On the other hand, anodic porous alumina templates are prepared by anodizing an aluminum foil. The cylindrical pores get arranged in a long-range hexagonal fashion with tunable pore diameters ranging from $8$ to more than $350\,$nm depending on the voltages and acids used during the anodization~\cite{Masuda97,Masuda98,Martin12,Manzano14,Resende19}. Moreover, such templates are mechanically and thermally stable. 

Although other methods like atomic layer deposition (ALD) have been used to fill the hollow templates described above~\cite{Pitzschel09}, the most common method found in literature to nanoengineer thermoelectric materials inside the templates has been electrodeposition (ED), so far. The main reason behind this trend is that the materials can be obtained near room temperature, the low fabrication cost, its high deposition rates, and the ability to tailor the properties of the materials by adjusting ED parameters. It is worth mentioning that the ED can be also used for growing NWs outside a template, for example using the step edges of highly oriented pyrolytic graphite as nucleation sites and coalescence to obtain Bi$_{2}$Te$_{3}$~\cite{Menke06}.

In order to electrodeposit inside the pores of the templates, for both types of templates, a metal layer (Ag, Au, Pt, Ni) is first deposited on one side of the membranes. This layer acts as a cathode during the electrochemical deposition, when immersed in an electroplating bath containing the semiconductor ions. A three-electrode electrochemical cell is generally used as the main deposition process when using these type of templates, due to its very high growth rates, great control, easy to up-scale to an industrial level, very simple apparatus and experimental procedure. 

By using both type of templates, different thermoelec\-trics have been prepared by electrodeposition~\cite{Boulanger03,Caballero-Calero16} from water solution or using organic solvents like dimethyl sulfoxide (DMSO), such as Bi~\cite{Liu98}, Bi$_{1\mathrm{-}x}$Sb$_{x}$~\cite{Prieto03, Keyani06},  Bi$_{2}$Te$_{3}$~\cite{Sapp99, Prieto01, Sander02, Sander03, Trahey07, Li06, Lee08, Wang06a, Peranio12, Manzano14, Frantz12}, Bi$_{2}$(TeSe)$_{3}$~\cite{MartinGonzalez03(a)}, Sb$_{2}$Te$_{3}$, PbTe, (BiSb)$_{2}$Te$_{3}$~\cite{MartinGonzalez03(b), Wang05}, PbSe~\cite{Klammer10}, PbSe$_{1\mathrm{-}}$\textit{$_{x}$}Te\textit{$_{x}$}, CoSb$_{3}$,  and Bi$_{2}$Te$_{3}$/Sb~\cite{Wang08} and Bi$_{2}$Te$_{3}$/(Bi$_{0.3}$Sb.$_{0.7}$)$_{2}$Te$_{3}$ superlattice NWs~\cite{Yoo07}. 

Regarding the fabrication of polymeric thermoelectric NWs, they have been also obtained by polymer infiltrations inside porous alumina, as in the case of P3HT, which showed a correlation of the thermal conductivity with the orientation of the polymer when confined to one-di\-men\-sion\-al structures~\cite{MunozRojo14,Taggart11}. Also, polymer NWs can be fabricated via lithographically patterned electrodeposition, like in PEDOT to obtain long NWs, which show enhanced properties compared to bulk which can be also associated to the crystallinity of the confined polymer. Finally, in order to study the properties or free-standing NWs, the templates can be dissolved selectively, using organic solvents in the case of ion-track membranes and using acid or bases in the case of porous alumina templates. Subsequently, the NWs can be obtained in solution for measurements in single entities. 
%
% The main results found in these kinds of single NWs will be explained in the properties section.

\begin{figure}[htb]
\centerline{\includegraphics[width=0.9\columnwidth]{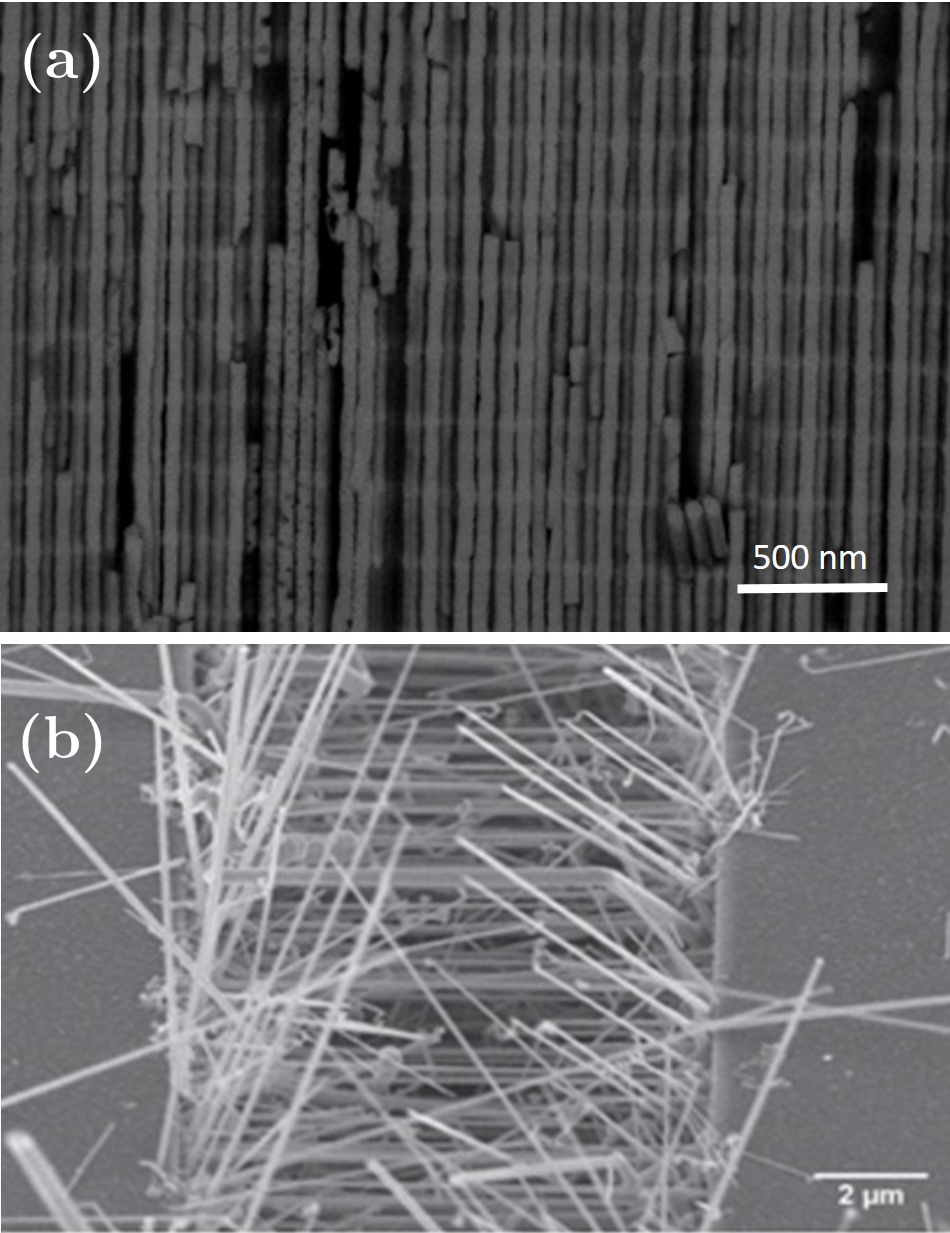}}
\caption{SEM micrograph of two examples of interconnection of NWs. (a)~Free-standing three-dimensional interconnected Bi$_2$Te$_3$ interconnected NWs~\cite{Caballero-Calero16, Martin14, Ruiz-Clavijo18} and integration of NWs into devices. (b)~Top view of the Si NW array obtained by means of bottom-up approach, connecting rim and suspended platform~\cite{Fonseca16}.}
\label{fig3}
\end{figure}

A step towards the integration of thermoelectric NWs in real bulk devices is the synthesis of three-dimensional interconnected NW networks, like the Bi$_{2}$Te$_{3}$ NWs networks shown in Fig.~\ref{fig3}, based on the electrodeposition inside three-dimensional porous anodic alumina templates developed in 2014~\cite{Martin14}, which consists of NWs connected periodically with each other by transverse channels of the same material, forming a kind of self-sustained scaffold that can hold together even when the matrix is dissolved. In these kinds of three-dimensional templates, thermoelectric materials have been grown via ED~\cite{Martin14, Ruiz-Clavijo18, Abad16} showing a reduction in their thermal conductivity due to the nano-structuration~\cite{Abad16}. Another approach for the integration of thermoelectric NWs in actual devices is the \emph{smart} integration of Si NW arrays in micro-nanogenerators~\cite{Davila12,Fonseca16}, also shown in Fig.~\ref{fig3}.%~\cite{Davila12}.

\subsection{Top-down thermoelectric nanowire fabrication}

The top-down approaches are mainly devoted to the fabrication of Si NWs. These NWs have been grown on Si wafers by different advanced lithographic tools, such as electron beam lithography~\cite{Ciucci05, Pennelli06}, atomic force lithography~\cite{Chien02, Legrand02, Martinez08}, and even optical litho\-graphy~\cite{Pott08, Pennelli09}. Etching, oxidation and other fabrication processes are then used to define structures with nanometric dimensions. As an example of the latest evolution in this field, the minimum feature size that can be achieved with these techniques is of less than $1\,$nm for scanning probe lithography~\cite{Gates04}, under $5\,$nm for direct-write charged particle lithography~\cite{Yang09}, around $10\,$nm for both nanoimprint litho\-graphy~\cite{Costner09} and optical litho\-graphy~\cite{Naulleau09}, and via interference litho\-graphy, features down to $15\,$nm can be achieved. 

Apart from mentioning the possibility of obtaining structures of a few nanometers, we will not enter in much detail in the top-down fabrication methods for the fabrication of thermoelectric NWs given that it is somewhat more complex compared to the bottom-up approach that allows for the massive, low-cost production of nanostructures. However, we should point out that the top-down fabrication approach allows for the simultaneous fabrication of nanostructures and NWs together with contacts, connections and control gates, so that the fabrication process yields fully functional devices, which is advantageous for some applications. Moreover, top-down fabrication techniques such as the fabrication of rough Si NWs by aqueous electroless etching presented by Hochbaum \emph{et al.}~\cite{Hochbaum08} demonstrated how the roughness in the NWs can decrease the thermal conductivity and thus, enhance the thermoelectric properties. 

Another interesting top down method to produce vertically aligned NWs is the so call metal-assisted chemical etching (MACE). In this method a few nanometer noble metal thin film is deposited on top of the thermoelectric material of interest. Then, this is immersed in an acid solution (normally HF) and an oxidant (normally H$_{2}$O$_{2}$). Under these conditions a redox reaction occur (electroless etching) and the initial process leads to the rupture of the metallic film into nanoparticles. While the process continues the nanoparticles etch away the semiconductor underneath generating tracks that leads to the formation of NWs in the areas where no nanoparticles were present. With this procedure vertical NWs of different thermoelectrics can be generated with typical length of $1-100\,\mu$m depending on the etching time~\cite{Huang11}. Nevertheless, generally speaking, top-down approaches are mainly used for Si and related materials, which are not the best candidates for thermoelectric applications.

As it has been shown through this section, there are several fabrication techniques that can be used to grow NWs and NW arrays using different approaches. Nevertheless, the final choice of the fabrication technique suitable for each application requires not only taking into account the ultimate resolution but other aspects, like the final configuration. As an example, anodic porous alumina and ion track-etched membranes are normally used when vertically aligned NWs are needed (out of plane configuration). Similarly, interference lithography is more suitable if the desired configuration is in the plane. Sometimes different techniques are used synergistically to obtain the desired result. As already seen in Table~\ref{table1}, all the approaches have benefits and drawbacks and the most adequate fabrication technique must be identified, depending on the final objective of the thermoelectric NW.

%%%%%%%%%%%%%%%%%%%%%%%%%%%%%%%%%%%%%%%%%%%%%%%%%%%%%%%%%%%
\section{Theoretical aspects}   \label{sec:th}
%%%%%%%%%%%%%%%%%%%%%%%%%%%%%%%%%%%%%%%%%%%%%%%%%%%%%%%%%%%

\subsection{Electron transport in nanowires}   

In this section we address the calculation of the electron contribution to charge and heat currents. In the linear response regime, carrier transport is driven by small driving forces, viz., electric field $\mathbf{E}$ and temperature gradient $-\nabla T$. Thus, the charge $\mathbf{i}$ and the heat $\mathbf{j}$ current densities along the NW axis obey the equations
\begin{align}\label{eq_linear}
\begin{bmatrix}
           \mathbf{i} \\
           \mathbf{j} 
\end{bmatrix} = 
\begin{bmatrix}
           {L}_{11} & L_{12} \\
           {L}_{21} & L_{22}
\end{bmatrix}
\begin{bmatrix}
           \mathbf{E} \\
           -\nabla T
\end{bmatrix}\,,
\end{align}
where the transport coefficients $L_{ij}$ form the Onsager matrix.

The semiclassical picture of electronic conduction assumes that electrons undergo scattering processes with an average collision time $\tau$. For a given background temperature $T$, electrons with energy $E$ are distributed according to the Fermi function $f(E)=1/\{1+\exp[(E-E_F)/k_B T]\}$ with $E_F$ the Fermi energy. At low $T$ the collision time can be evaluated at $E_F$ and the contribution to transport arises from states around $E_F$. Hence, the electric conductivity is given by~\cite{Ashcroft76}
\begin{equation}\label{eq_L11}
L_{11} = \sigma = \frac{e^2\tau}{2\pi}\int D(E)f(E) m^{-1}(E) \,dE \,.
\end{equation}
Here, $D(E)$ is the density of states and $m$ is the effective mass, which will henceforth be assumed to be independent of energy. The one-band model considers a single confined state in the transverse directions of the NW with energy $\varepsilon_0$. In the transport direction, electrons are free, which amounts to having a parabolic conduction subband. Then, $D(E)$ is proportional to $\theta(E-\varepsilon_0)(E-\varepsilon_0)^{-1/2}$ with $\theta(x)$ the Heaviside function. The divergence of the density of states as $E\to \varepsilon_0$ is characteristic of one-dimensional systems (van Hove singularity). Substituting in Eq.~\eqref{eq_L11} one finds~\cite{Hicks93b}
\begin{equation}\label{eq_cond}
\sigma = \frac{e^2 \tau}{\pi \hbar a^2} \left(\frac{2 k_B T}{m}\right)^{1/2} F_{-1/2}\,,
\end{equation}
where
\begin{equation}\label{eq_Fn}
F_\gamma(\eta_F) =  \int_0^\infty \frac{x^{\gamma}}{1+\exp(x-\eta_F)} \,dx
\end{equation}
is the Fermi-Dirac integral of order $\gamma$, as defined in Eq.~(6) of Ref.~\cite{Hicks93b}, which depends on $\eta_F = (E_F-E_0)/k_B T$, and $a$ is the NW width. Importantly, the lower limit of the integral in Eq.~\eqref{eq_Fn} has been shifted so that energies are evaluated from the subband bottom.

A similar calculation shows that the Seebeck coefficient
\begin{equation}
    S = \frac{k_B}{e} \left(\eta_F - \frac{3F_{1/2}}{F_{-1/2}} \right) \,,
\end{equation}
is independent of $a$. On the other hand, the functional dependence of $\kappa_\mathrm{el}$ is $a^{-2}$ like the electric conductivity given by Eq.~\eqref{eq_cond},
\begin{equation}
    \kappa_\mathrm{el} = \frac{2\tau k_B^2T}{\pi a^2}\left(\frac{2k_BT}{m\hbar^2}\right)^{1/2}
    \left(\frac{5}{2}\,F_{3/2} - \frac{9F_{1/2}^2}{2F_{-1/2}} \right) \,,
\end{equation}
where $\tau$ is the relaxation time. As a consequence, the figure of merit increases for decreasing $a$. Ideally, a narrow NW would be an excellent thermoelectric generator and cooler. Nevertheless, in the limit $a\to 0$ quantum effects are unavoidable as we now discuss.

When the NW is free of disorder, transport becomes ballistic and the semiclassical theory breaks down. A full quantum-mechanical calculation is then in order. Here, we briefly discuss the Keldysh-Green function approach, a powerful formalism that can treat both non-equilibrium situations and electron-electron interactions. The key object is the retarded Green function, $G^r(E)$. Quite generally, its analytic structure provide us with information about the excitation energies of the system and their lifetimes. The latter are encoded in the self-energies $\Sigma_\alpha(E)$ labeled with a lead index (e.g., $\alpha = \{L,R\}$ for a NW connected to two terminals, left and right). NW electronic states decay into the the terminals due to tunnel effect. Therefore, the quantum mechanical transmission probability that an an electron traverses the NW is given by the following expression~\cite{Meir92}
\begin{equation}\label{eq_tr}
T(E) = {\rm Tr} \big\{ \Gamma_L (E) G^r (E) \Gamma_R (E) \left[G^r (E)\right]^\dagger\big\}\,,
\end{equation}
where $\Gamma_\alpha(E)=-2\,{\rm Im}\,\Sigma_\alpha(E)$ measures the coupling between the NW and the two leads at an energy $E$. In mesoscopic transport, $T(E)$ determines the conduction properties as shall be seen now. Within the Landauer-B\"uttiker formalism, the leads are considered to be mas\-sive fermionic reservoirs in local thermodynamic equilibrium with well defined electrochemical potentials $\mu_\alpha=E_F+eV_\alpha$ and $T_\alpha=T+\theta_\alpha$ ($V_\alpha$ and $\theta_\alpha$ are the voltage and temperature biases, respectively). Their Fermi distribution functions should account for this: $f_\alpha(E)=1/\{1+\exp[(E-\mu_\alpha)/k_B T_\alpha]\}$. Then, the electric and heat currents through the NW and measured at, say, the left lead are expressed as
\begin{align}\label{eq_I}
I &= \frac{e}{h}\int_0^\infty T(E) [f_L(E)-f_R(E)] \,dE \,, \\
J &= \frac{1}{h}\int_0^\infty (E-\mu_L) T(E) [f_L(E)-f_R(E)] \,dE \label{eq_J}\,.
\end{align}
To compare with Eq.~\eqref{eq_linear} we expand Eqs.~\eqref{eq_I} and~\eqref{eq_J} up to first order in $V_\alpha$ and $\theta_\alpha$, yielding 
\cite{Butcher90}
\begin{align}
\begin{bmatrix}
           I \\
           J 
\end{bmatrix} = 
\begin{bmatrix}
           G & L \\
           M & K
\end{bmatrix}
\begin{bmatrix}
           V \\
           \theta
\end{bmatrix}\,.
\end{align}
with $V=V_L - V_R$, $\theta = \theta_L - \theta_R$ and
\begin{subequations}
\begin{align}
G &= \frac{e^2}{h} \int T(E) \left(-f^{\prime}\right)_{\rm eq} \,dE \,, \label{eq_G} \\
L &= \frac{M}{T} = \frac{e^2}{hT} \int (E-E_F) T(E) (-f^{\prime})_{\rm eq} \,dE \,, \label{eq_rec}\\
K &= \int (E-E_F)^2 T(E) (-f^{\prime})_{\rm eq} \,dE\,.
\end{align}
\end{subequations}
Here, the subscript \textquotedblleft eq\textquotedblright means equilibrium ($V=\theta=0$) and $f^{\prime}=\partial f/\partial E$. The transmission function given by Eq.~\eqref{eq_tr} is also calculated at equilibrium. The first equality of Eq.~\eqref{eq_rec} is ensured by reciprocity between charge and heat transport in linear response. For extensions into the nonlinear regime of transport, we refer the reader to the review of Ref.~\cite{Sanchez16}.

We now need to assess the Green function. For independent electrons, $G^r(E)$ reduces to the resolvent of the system Hamiltonian, $G^r(E) = (E - H +i0^+)^{-1}$, where $H=H_{l}+H_{w}+H_t$ is split into the Hamiltonian for the leads $H_{l}$, the Hamiltonian for the NW $H_{w}$, and the tunnel term that couples both subsystems $H_t$. For definiteness, we consider that the wire is a hollow tube of radius $R$. In polar coordinates the NW Hamiltonian reads
\begin{equation}\label{eq_ham}
 H_{w}=\frac{-\hbar^2}{2m} \left[\partial_x^2+\frac{1}{R^2}\,\frac{\partial^2}{\partial \varphi^2}\right] \,.
\label{eq:Hamiltonian}
\end{equation}
The case with applied magnetic fields can be treated with an appropriate change of the momentum operator~\cite{Erlingsson17}. The eigenstates of Eq.~\eqref{eq:Hamiltonian} are $\psi= e^{i\varphi n} e^{i k x}$ with $n= 0, \pm 1, \pm 2\ldots$ and $k$ the wavevector. These states represent traveling waves along $x$ and their energies, $E= \hbar^2 k^2/2m + \varepsilon_n$ with $\varepsilon_n=\hbar^2 n^2/2mR^2$, are described with parabolas vertically shifted as $|n|$ (basically, the angular momentum) increases. If the NW cross section is a square of width $a$ along the $y$ and $z$ directions, then the eigenergies are replaced with $E= \hbar^2 k^2/2m + \hbar^2 n_y^2/2ma^2 + \hbar^2 n_z^2/2ma^2$, where $n_y,n_z=1,2,\ldots$ For more complex geometries, the Hamiltonian should be numerically diagonalized. For this purpose, a convenient approach is the quantum transmitting boundary method, which can deal with small conductors of arbitrary shape and couplings~\cite{Lent90} and has been recently applied to thermoelectric transport~\cite{Sanchez11,SaizBretin15}.

For a fully transparent NW the energy dispersion determines the thermopower (Seebeck coefficient) since the transmission is simply unity if $E>\varepsilon_n$ and 0 otherwise. At zero temperature, the conductance given by Eq.~\eqref{eq_G} becomes quantized as the energy increases, showing plateaus at multiples of $e^2/h$~\cite{vanweperen13,kammhuber16}. The situation is akin to a quantum point contact~\cite{vanwees88,wharam88}, where the constriction in a two-dimensional electron gas plays the role of a quantum wire. The thermopower of quantum point contacts has been studied both theoretically~\cite{streda89,proetto91} and experimentally~\cite{molenkamp90,dzurak93}. A salient feature is the oscillations of $S$ as a function of the Fermi energy by gate modulation. This phenomenon can be understood from the Cuttler-Mott relation~\cite{cutler69,lunde05}, which at low temperatures establishes a connection between the thermopower and the logarithmic derivative of the conductance
\begin{equation}
    S = -\frac{\pi^2 k_B^2 T}{3e} \frac{1}{G}\frac{dG}{dE_F} \,.
\end{equation}
In those regions where $G$ shows plateaus $S$ is zero whereas in the transitions between plateaus the conductance increases, which causes peaks in the thermopower that are periodically spaced with the number of open channels. This oscillatory behavior has also been observed in InAs NWs with large subband splittings~\cite{tian12}. Interestingly, NWs serve as tools to test the quantum bound of thermoelectric power output~\cite{chen18}. It should be also noted that quantum dots can be formed inside NWs by epitaxially growing double tunnel barriers with semiconductor heterostructures. Then, the local density of states effectively becomes zero dimensional and the physics and thermoelectric properties accordingly change: Coulomb blockade effects due to charging effects~\cite{svensson13}, interference induced power factor enhancements~\cite{wu13}, thermally assisted Kondo effects~\cite{svilans18}, just to mention a few.

We finish this discussion with a brief theoretical analysis of topological NWs. These systems are similar to core-shell NWs [see Eq.~\eqref{eq_ham}] because both conduct electrons on the surface while the bulk remains insulating. However, carriers in topological (core-shell) NWs are Dirac (Schr\"odinger) fermions, which lead to metallic states protected from disorder as discussed in Sec.~\ref{sec:tinw}. This topology boundary can be understood by means of a linear-in-momentum Hamiltonian~\eqref{eq_ham}~\cite{bardarson13}:
\begin{equation}\label{eq_hamti}
    H_w = -i\hbar v_F \left(\sigma_z \partial_z + \sigma_y \frac{1}{R}\partial_\varphi \right)\,.
\end{equation}
Here, $\sigma_{y,z}$ are Pauli matrices and $v_F$ is an interband matrix element having dimensions of velocity that can be assumed scalar. As such, $v_F$ is material dependent. The eigenstates of $H_w$ are spinors with an extra $\pi$ factor due to the Berry phase that leads to a minimum conductance
at zero magnetic field. The Hamiltonian given by Eq.~\eqref{eq_hamti} is an effective model within the envelope-function approximation. The energy bands in topological NWs also exhibit interesting differences. In contrast to Schr\"odinger NWs, topological spectra are not bounded from below. Further, the gap $E_G$ between positive and negative energies can be closed (Dirac point) upon application of a magnetic field. Hence, conductance becomes dominated by a perfectly conducting channel without backscattering, similarly to metallic carbon nanotubes~\cite{ando02}.

\begin{figure}[htb]
\centerline{\includegraphics[width=0.8\columnwidth]{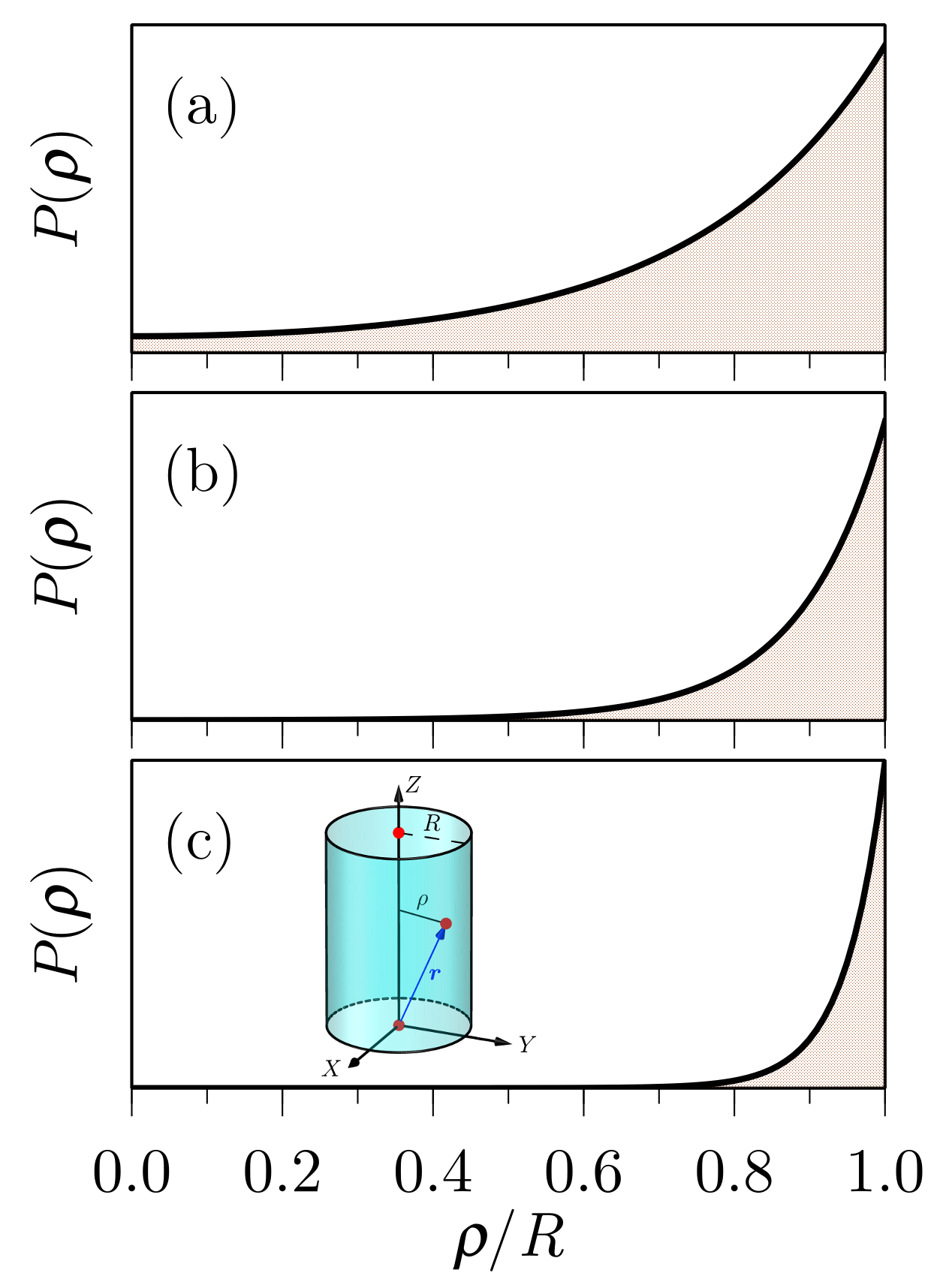}}
\caption{Radial probability density, in arbitrary units, of the mid-gap surface states for Bi$_2$Te$_3$ NWs of radius (a)~$R=\SI{20}{\nano\meter}$, (b)~$R=\SI{40}{\nano\meter}$ and (c)~$R=\SI{80}{\nano\meter}$. In all cases the total angular momentum quantum number is $j=1/2$. Data courtesy of Jorge Estrada.}
\label{fig4}
\end{figure}

If we take into account the spin degree of freedom, Eq.~\eqref{eq_hamti} becomes a $4\times 4$ matrix~\cite{Agassi88,Pankratov90,Hsieh12,Adame94,Kolesnikov97,Kane12,Tchoumakov17}:
\begin{equation}\label{eq_hamti2}
H_w=-i\hbar v_F\,{\bm\alpha}\cdot{\bm{\nabla}}+\frac{1}{2}\,E_{G}(\rho)\,\beta\ .
\label{eq:05}
\end{equation}
where ${\bm\alpha}=(\alpha_x,\alpha_y,\alpha_z)$ and $\beta$ denote the usual Dirac matrices. The energy gap $E_{G}(\rho)$ depends on the radial coordinate $\rho$ and takes on two values, namely, $E_{G,\mathrm{NW}}$ for $0\leq \rho\leq R$ and infinite otherwise. We note that Eq.~\eqref{eq_hamti2} acts upon the envelope function ${\bm F}({\bm r})$, which is a bispinor whose spinor components belong to two bands. We assume that $H_w$ possesses a translational symmetry along the direction of the NW, which implies that the axial momentum $k_x$ is a good quantum number. After solving the corresponding eigenenergy equation $\mathcal{H}{\bm F}({\bm r})=E{\bm F}({\bm r})$, a set of mid-gap states localized at the vicinity of the surface is found. The probability density $P={\bm F}^{\dag}\cdot{\bm F}$ depends only on the radial coordinate $\rho$. Fig.~\ref{fig4} shows $P(\rho)$ for surface states with total angular momentum quantum number $j=1/2$, namely the deepest state in the gap, for three different values of the radius of Bi$_2$Te$_3$ NWs. It becomes apparent that surface states extend noticeably towards the center of the NW, especially in narrow ones. States with larger values of $j$ have an energy closer to the band edges but their spatial extend is similar to the states with $j=1/2$. The radial extent is given as $2\hbar v_F/E_{G,\mathrm{NW}}$, being of the order of $\SI{4.1}{\nano\meter}$ in Bi$_2$Te$_3$ and $\SI{2.8}{\nano\meter}$ in Bi$_2$Se$_3$~\cite{Madelung98,Diaz-Fernandez18b}. 
\reivision{It is worth mentioning that some authors claim that the contribution of electron surface states to thermoelectric properties are relevant only in narrow nanostructures, whilst the observed results in wider ones can be explained by bulk band bending~\cite{Thompson13}. However, there is evidence that this is not necessarily so in all cases~\cite{MunozRojo16}.}

Topologically protected surface states are robust under perturbations such as defects and surface roughness. In this regard, several works have demonstrated that surface roughness is harmless for the electric conduction in narrow NWs~\cite{MunozRojo16,Kong10}. Most importantly, the lattice contribution to the thermal conductivity is reduced by the presence of the surface. For instance, non-equilibrium molecular dynamics simulations point out a noticeable reduction of $\kappa_\mathrm{ph}$ in Bi$_2$Te$_3$ NW~\cite{Yu14}. Therefore, topological insulator NWs holds all the characteristics to be the building blocks of highly efficient thermoelectric devices. It is worth mentioning that there is a single key parameter characterizing the thermoelectric performance of topological insulator NWs, namely the surface-to-volume ratio. The thermoelectric coefficients change significantly with increasing this ratio, as it has been highlighted by several works~\cite{Shin16,Kong10,Gooth15,MunozRojo16}. 

\subsection{Lattice thermal conductivity}

The lattice thermal conductivity in semiconductors and insulators is usually the main contribution to the thermal conductivity of the material. For instance, Hernandez \emph{et al.} reported that the contribution due to free charges is in some cases up to two orders of magnitude lower than the contribution due to phonons in SnSe NWs~\cite{Hernandez18}. In this section we briefly discuss a number of theoretical approaches that are routinely used in the calculation of $\kappa_\mathrm{ph}$. 

\subsubsection{Non-equilibrium molecular dynamics}

When the system is subjected to a temperature gradient $\nabla T$, the heat current density ${\bm j}_\mathrm{Q}$ due to the atomic lattice satisfies the Fourier law
\begin{equation} 
{\bm j}_\mathrm{Q}=-\kappa_\mathrm{ph} \nabla T\,.
\label{Fourier}
\end{equation}
Non-equilibrium molecular dynamics (NEMD) simulations bring a straightforward method to calculate $\kappa_\mathrm{ph}$ by applying a temperature gradient to the system and measuring the resulting heat current density~\cite{Plimpton95}. In this approach, the trajectories of classical point particles (atoms) of mass $M$ are obtained by numerically solving Newton's equations of motion. Atoms with initial positions and velocities are subjected to collisions governed by an empirical interatomic potential. At every time step, the force acting on each atom is obtained and then positions and velocities are updated~\cite{Frenkel96}. This is an excellent approximation for a wide range of materials except when one considers light atoms or vibrational motion with frequency larger than $k_BT/\hbar$. In these cases, quantum effects become important~\cite{Frenkel96}. At lower temperatures, quantum corrections have been developed to mitigate this limitation~\cite{Hu09}. In general, the classical treatment is valid whenever the mean interparticle distance is much larger than the thermal de Broglie wavelength $h/\big(2\pi M k_B T\big)^{1/2}$. 

The small size of a nanoscale system poses a question about the validity of Fourier's law~\eqref{Fourier}. In this regard, Kaiser \emph{et al.} presented a thorough study on steady-state thermal transport in the nanoscale and concluded that Fourier's law is essentially exact in the diffusive and ballistic limits (see Ref.~\cite{Kaiser17} and references therein for further details).

Wang \emph{et al.} have calculated the phonon thermal conductivity of Si NWs using NEMD simulations with the empirical Stillinger-Weber potential model~\cite{Stillinger85} and the Nos\'{e}-Hoover thermostat~\cite{Nose84,Hoover85}. The thermal conductivity dependence on temperature was found to agree well with experiments over a wide range of cross-sections and lengths~\cite{Wang09b}. Applying NEMD simulations, it was demonstrated that the thermal conductivity of Si NWs can be reduced remarkably by shell doping due to both impurity scattering and interface scattering associated with the peculiar structure of shell-doped geometries~\cite{Wang13}. Similar effects were observed in Si NWs with faulted stacking layers~\cite{Zhan14}, porous Si NWs~\cite{Cartoixa16} and He-irradiated Si NWs~\cite{Zhao17}. Despite the general trend of an expected decrease of the thermal conductivity upon reducing the diameter of the NW, NEMD simulations with the Stillinger–Weber potential model of very narrow Si NWs (diameters smaller than $\SI{1.5}{\nano\meter}$) predict an increase of this magnitude~\cite{Ponomareva07}. This enhancement has been attributed to phonon confinement effects (see also Ref.~\cite{Donadio09}). Besides Si NWs, NEMD simulations were used to explore other systems, such as diamond NWs~\cite{Guo10}.

\subsubsection{Phonon Boltzmann transport equation} 

The semiclassical description of phonon transport based on the Boltzmann equation provides a computationally  af\-ford\-able approach to calculate $\kappa_\mathrm{ph}$ for experimentally relevant sizes. Let ${\bm v}_{j}({\bm q})$ be the phonon group velocity, where ${\bm q}$ is the phonon wavevector and the subscript $j$ refers to a particular phonon branch. The heat current density can be then obtained as~\cite{Ibach09,Zou01}
\begin{equation} 
{\bm j}_\mathrm{Q}=\frac{1}{V}\sum_{{\bm q},j}\hbar \omega _{j}({\bm q})
\Big[ \langle n_{j}({\bm q})\rangle -
\langle n_{j}({\bm q})\rangle_{\mathrm{eq}} \Big]{\bm v}_{j}({\bm q})\,,
\label{heat_current_density}
\end{equation}
where $V$ is the volume of the system, $\hbar \omega _{j}({\bm q})$ is the phonon energy and $\langle n_j({\bm q})\rangle$ is the phonon distribution. In equilibrium it is given by the Bose-Einstein distribution function
\begin{equation} 
\langle n_{j}({\bm q})\rangle_{\mathrm{eq}} 
=\frac{1}{\exp \big[\hbar\omega _{j}({\bm q})/k_{B}T\big]-1}\,.
\end{equation}
In steady state, the phonon Boltzmann transport equation is written as follows
\begin{equation} 
\frac{\partial \langle n_{j}({\bm q})\rangle}{\partial t} \Bigg|_\mathrm{drift}
+\frac{\partial \langle n_{j}({\bm q})\rangle}{\partial t} \Bigg|_\mathrm{coll}
=0 \,,
\label{Boltzmann1}
\end{equation}
The first term represent the change of the phonon distribution due to the presence of the temperature gradient. Assuming that the spatial variation of the temperature is small, it is found to be~\cite{Ibach09,Zou01}
\begin{equation} 
\frac{\partial \langle n_{j}({\bm q})\rangle }{\partial t} \Bigg|_{\mathrm{drift}}
= {\bm v}_{j}({\bm q})\cdot\nabla T\,
\frac{\partial \langle n_{j}({\bm q})\rangle_{\mathrm{eq}}}{\partial T}\,.
\medskip
\end{equation}
The second term in Eq.~\eqref{Boltzmann1} takes into account the changes of the phonon distribution due to scattering by other pho\-nons, charge carriers, defects and system boundaries. In the relaxation-time approximation it is given by
\begin{equation} 
\frac{\partial \langle n_{j}({\bm q})\rangle }{\partial t}\Bigg|_\mathrm{coll}
=-\frac{ \langle n_{j}({\bm q})\rangle 
-\langle n_{j}({\bm q})\rangle_{\mathrm{eq}}}{\tau _{j}({\bm q})} \,.
\end{equation}
$\tau _{j}({\bm q})$ being the relaxation time. Collecting all terms, the phonon Bolztmann equation~\eqref{Boltzmann1} turns out to be
\begin{equation} 
{\bm v}_{j}({\bm q})\cdot\nabla T\,
\frac{\partial \langle n_{j}({\bm q})\rangle_{\mathrm{eq}}}{\partial T}
=\frac{ \langle n_{j}({\bm q})\rangle 
-\langle n_{j}({\bm q})\rangle_{\mathrm{eq}}}{\tau _{j}({\bm q})} \,.
\label{Boltzmann2}
\end{equation}
Once the phonon Boltzmann transport equation is solved, the contribution of the lattice to the thermal conductivity is calculated with the aid of Eqs.~\eqref{Fourier} and~\eqref{heat_current_density}. For instance, in bulk one obtains the well-known Klemens-Callaway's expression for $\kappa_\mathrm{ph}$ (see Ref.~\cite{Srivastava90} and references therein). 

Zou and Balandin~\cite{Zou01} used the phonon Boltzmann transport equation~\eqref{Boltzmann2} to study lattice heat conduction in a Si NW with dimensions comparable to $\ell_\mathrm{ph}$, including the modification of $\omega_j({\bm q})$ due to spatial confinement and changes in $\langle n_{j}({\bm q})\rangle$ due to boundary scattering from the rough surface. Both effects lead to a significant decrease of $\kappa_\mathrm{ph}$. Similar conclusions were drawn by Abouelaoualim regarding Bi$_{0.95}$Sb$_{0.05}$ NWs~\cite{Abouelaoualim07}. Huang \emph{et al.} studied theoretically Si NWs and derived analytically an expression of the boundary scattering rate for phonon heat transport~\cite{Huang06}. In addition to the general trait of a marked reduction of $\kappa_\mathrm{ph}$ upon decreasing the NW diameter, they also found that the lattice contribution to the thermal conductivity decreases after reducing the fraction of specularly reflected phonons at the surface. In Ref.~\cite{Yang05}, Yang \emph{et al.} reported the dependence of $\kappa_\mathrm{ph}$ on the surface conditions and the core-−shell geometry for Si core-−Ge shell and tubular Si NWs at room temperature. Their results show that the effective thermal conductivity changes not only with the composition of the constituents but also with the radius of the NW and nanopores.

\subsubsection{First principles calculations}

The reliability of NEMD simulations is generally affected by the choice of the phenomenological interatomic potential for the particles constituting the system. In other words, the employed potential needs to be sufficiently representative of the actual atomic interactions. Once the potential is set, NEMD simulations can handle a large number of atoms with modest computational resources. First principles methods, such as density functional theory (DFT), GW approximation or Bethe-Salpeter equation, are based on the laws of quantum mechanics and only need the fundamental physical constants as input, without invoking phenomenological fitting parameters. Although the number of constituent atoms in a simulation is limited, first principles methods provide a detailed insight into the interatomic potential that can be used in combination with NEMD simulations or phonon Boltzmann transport equation. 

Combining equilibrium molecular dynamics, DFT and Boltzmann transport theory methods, Demchenko \emph{et al.} provided a complete picture for the competing factors of $ZT$ in ZnO and Si NWs, finding that $ZT$ can be increased as much as $30$ times in $\SI{0.8}{\nano\meter}$-diameter ZnO NWs and $20$ times in $\SI{1.2}{\nano\meter}$-diameter Si NWs, compared with the bulk~\cite {Demchenko11}. By decoupling the lattice and electron contributions to the thermal conductivity, they claimed that the improvement of the thermoelectric response was mainly due to the reduction of $\kappa_\mathrm{ph}$. Using a combination of DFT and Boltzmann transport calculations, Song \emph{et al.} studied Bi$_2$Te$_3$ NWs and concluded that the electronic thermal transport can be decoupled from the electrical conductivity by changing the density of surface states~\cite{Song18}. The same approach was recently employed by Akiyama \emph{et al.} to propose a new way to design high-performance thermoelectric materials by controlling the polytype of Si and Ge NWs~\cite{Akiyama17}. Similarly, using the same approach, Lu \emph{et al.} demonstrated that Bi NWs present a largely enhanced $ZT$, of the order of $2.73$ at room temperature, due to a significant reduction of $\kappa_\mathrm{ph}$~\cite{Lu17}. The decline of the lattice thermal conductivity was attributed to the reduced phonon vibration frequency, the decreased phonon density of states and the shortened $\ell_\mathrm{ph}$. 

%%%%%%%%%%%%%%%%%%%%%%%%%%%%%%%%%%%%%%%%%%%%%%%%%%%%%%%%%%%
\section{Applications}   \label{sec:app}
%%%%%%%%%%%%%%%%%%%%%%%%%%%%%%%%%%%%%%%%%%%%%%%%%%%%%%%%%%%

Thermoelectric materials are useful in power generation and refrigeration devices without moving parts. As already discussed above, it is now well established that thermoelectric properties can be largely improved by structuring it into arrays of NWs and carefully controlling their morphology and doping~\cite{Boukai08}. Besides fundamental knowledge on the interplay between electric and thermal transport, applications are a driving force in thermoelectric research as well~\cite{He15,Champier17}. 

NWs find a natural niche for applications in electronics as micro- and nano-generators due to a lower energy requirements (e.g. intelligent autonomous sensors)~\cite{Gadea18}. In particular, Si NWs present a reasonable balance between technological feasibility in current devices and fabrication techniques on one side and high thermoelectric efficiency compared to bulk Si on the other side~\cite{Pennelli14}. In Ref.~[\citenum{Fonseca16}]  Fonseca~\emph{et al.} presented two approaches for the integration of large numbers of Si NWs in a cost-effective way using only micromachining and thin-film processes compatible with Si technologies. Both approaches enable automated integration of stacked arrays of Si NWs with diameters below $\SI{100}{\nano\meter}$ for thermoelectric micro-nanogenerators. Similarly, Totaro~\emph{et al.} devised a top down process for the fabrication of nets of well organized and connected Si NWs, finding high reliability with respect to NW failure~\cite{Totaro12}. These nets of NWs are suited for high efficiency thermoelectric generators or for nanosensors. Fabrication and characterization of single, double, and quadruple stacked flexible Si NW network based thermoelectric modules have been reported by Norris \emph{et al.}~\cite{Norris15}. Remarkably, power production increased $27\%$ from double to quadruple stacked modules, demonstrating that stacking multiple NW thermoelectric devices in series is a scalable method to generate power by supplying larger temperature gradient. Field-effect modulation of the thermoelectric response in multigated Si NWs allow the electric conductivity and Seebeck coefficient to be tuned, as demonstrated by Curtin~\emph{et al.}~\cite{Curtin13}. The reported power factor is similar to the highest values for $n$-type Si nanostructures. Potentialities of Si NW forest for thermoelectric generation have been recently discussed by Dimaggio and Pennelli in Ref.~[\citenum{Dimaggio18}], where they provide general criteria for improving energy conversion. Finally, let us mention that materials other than Si are currently under intense research as well. As an example, Choi~\emph{et al.}~\cite{Choi14} have proposed a new system based on Te NW films hybridized with single-walled C nanotubes. The excellent mechanical stability and electrical conductivity of the nanotubes enhance the flexibility and thermoelectric properties of the pure Te NW film. These results indicate that the hybrid film would be promising for a potential use as a flexible thermoelectric materials.

%%%%%%%%%%%%%%%%%%%%%%%%%%%%%%%%%%%%%%%%%%%%%%%%%%%%%%%%%%%
\section{Conclusions and prospective} \label{sec:con}
%%%%%%%%%%%%%%%%%%%%%%%%%%%%%%%%%%%%%%%%%%%%%%%%%%%%%%%%%%%

A deep-rooted aim of thermoelectric research is concerned with improving the efficiency of the interconversion of heat and electricity in various ways. The quest for new and innovative material systems with increased heat-to-electricity conversion capabilities is one of the current goals in this active field. On the other side, recent advances in nanotechnology have brought with them promising methods that exploits fundamental mechanisms to achieve better thermoelectric devices. The proposal of reducing the dimensionality to significantly improve the figure of merit about the best values in bulk materials trace back to the pioneering work by Hicks and Dresselhaus \cite{Hicks93a,Hicks93b}. In particular, it was then proposed that a one-dimensional conductor whose width is narrower than the thermal de Broglie wavelength might display an enhanced figure of merit~\cite{Hicks93a}. In this context, semiconductor NWs arise as ideal candidates that meet these requirements. 

The present article comprises an overview of some of the fundamental attainments and breakthroughs in experimental and theoretical thermoelectric research based on NWs. They can manifest values of the figure of merit two order of magnitudes higher relative to their corresponding bulk counterparts, approaching the threshold ($ZT\sim 3$) for extensive applications. After intensive research, a reasonable understanding of the factors that promote their unusually high efficiency has been reached. Independent measurements of the Seebeck coefficient, electrical conductivity and thermal conductivity have led to the conclusion that limited phonon transport by surface scattering is responsible of the observed enhanced efficiency. NWs made of topological insulators represent a step forward since their surface electron states are protected by symmetries that reduce or even suppress backscattering, preserving electron transport. In spite of all these advances, further experimental and theoretical studies will turn NWs into a noteworthy route toward efficient thermoelectric devices.

%%%%%%%%%%%%%%%%%%%%%%%%%%%%%%%%%%%%%%%%%%%%%%%%%%%%%%%%%%%
\section*{Acknowledgment}
%%%%%%%%%%%%%%%%%%%%%%%%%%%%%%%%%%%%%%%%%%%%%%%%%%%%%%%%%%%

This work was supported by the Spanish MICINN under grants  MAT2016-63955-R, MAT2016-75955, MAT2017-82639 and MAT2017-86450-C4-3-R. The authors are indebted to O. Caballero, A.~D\'{\i}az-Fern\'{a}ndez, J. Estrada, R. Mart{\'i}nez and M.~Saiz-Bret\'{\i}n for enlightening discussions.

\appendix

%%%%%%%%%%%%%%%%%%%%%%%%%%%%%%%%%%%%%%%%%%%%%%%%%%%%%%%%%%%
\section{List of symbols}
%%%%%%%%%%%%%%%%%%%%%%%%%%%%%%%%%%%%%%%%%%%%%%%%%%%%%%%%%%%

For convenience, symbols used in the text are listed alphabetically below. 

\begin{table}[ht]
\setlength{\tabcolsep}{0.5em}     % for the horizontal padding
\renewcommand{\arraystretch}{1.2} % for the vertical padding
\small
\centering
\begin{tabular}{l|l}
\toprule 
Symbol & Description \\ 
\midrule
$C_\mathrm{v}$ & Specific capacity of phonons \\
               & at constant volume \\
$D$ & Density of states \\
$e$ & Elementary electric charge\\
$E_F$ & Fermi energy \\
$E_G$ & Energy gap \\
$f$ & Fermi function \\
${\bm F}$ & $4$-component envelope function \\
$G$ & Electric conductance \\
$G^r$ & Retarded Green's function \\
$H$ & Hamiltonian operator \\
$k$ & Wavevector \\
$k_B$ & Boltzmann constant \\
$\ell_\mathrm{el}$ &  Electron mean free path \\
$\ell_\mathrm{ph}$ &  Phonon mean free path \\
$L$ & Lorenz number \\
$m$ & Electron effective mass \\
$\langle n_j\rangle $ & Phonon distribution in the branch $j$\\
$P$ & Probability density of surface states\\
${\bm q}$ & Phonon wavevector\\
$R$ & Nanowire radius \\
$S$ & Seebeck coefficient\\
$T$ & Absolute temperature \\
$v$ & Acoustic phonon velocity \\
$v_F$ & Interband matrix element \\
$ZT$ & Figure of merit \\
$\alpha_x$, $\alpha_y$, $\alpha_z$, $\beta$ & $4\times 4$ Dirac matrices \\
$\kappa$ & Thermal conductivity \\
$\kappa_\mathrm{el}$ & Electron contribution to $\kappa$\\
$\kappa_\mathrm{ph}$ & Lattice contribution to $\kappa$\\
$\sigma$ & Electric conductivity \\
$\sigma_x$, $\sigma_y$, $\sigma_z$ & $2\times 2$ Pauli matrices \\
$\omega_j$ & Phonon frequency in the branch $j$\\
\bottomrule
\end{tabular}
\end{table}

%%%%%%%%%%%%%%%%%%%%%%%%%%%%%%%%%%%%%%%%%%%%%%%%%%%%%%%%%%%
\section*{References}
%%%%%%%%%%%%%%%%%%%%%%%%%%%%%%%%%%%%%%%%%%%%%%%%%%%%%%%%%%%

\bibliography{references}

\bibliographystyle{model1a-num-names.bst}

\end{document}